# Selective and enhanced nickel adsorption from sulfate- and calcium-rich solutions using chitosan

Nina Ricci Nicomel[a,*], Lila Otero-Gonzalez[a,I], Karel Folens[b], Bernd Mees[a], Tom Hennebel[b], Gijs Du Laing[a]

[a] Laboratory of Analytical Chemistry and Applied Ecochemistry, Department of Green Chemistry and Technology, Ghent University, Coupure Links 653, 9000 Ghent, Belgium

[b] Center for Microbial Ecology and Technology (CMET), Department of Biotechnology, Ghent University, Coupure Links 653, 9000 Ghent, Belgium

* Corresponding author:

E-mail: NinaRicci.Nicomel@UGent.be

Tel.: +3292646131

Fax: +3292646232

Present address:

[I] IDENER, Earle Ovington 24, 8-9, 41300 La Rinconada, Seville, Spain

**Highlights**

- Chitosan effectively adsorbed Ni from a real Ni/Ca/$SO_4^{2-}$ bearing leachate

- Sulfate had a positive effect on Ni adsorption while Ca did not have any effect

- Formation of $NiSO_4^0$ species contributed to the enhanced Ni adsorption onto chitosan

- Chitosan has a Ni $q_{max}$ of 1.49 mmol/g even with 500 mM $SO_4^{2-}$ and 10 mM Ca present

- Chitosan was selective for Ni over Ca as validated in a column setup

**ABSTRACT**

Nickel (Ni) is an economically important metal characterized by its mechanical strength and anticorrosion properties. With the increasing industrial demand for Ni and the depleting accessible Ni primary ores, sustainable technologies are required for the recovery of this metal from alternative resources. In this study, adsorption using chitosan was investigated as a sustainable technique to recover Ni from sulfate ($SO_4^{2-}$) and calcium (Ca) rich secondary resources. The effects of pH, contact time, and the presence of $SO_4^{2-}$ and Ca on Ni adsorption were investigated in batch experiments. Chemical speciation modeling was performed to analyze how the predominant Ni species present under different conditions may affect the efficiency of the adsorption process. The comparison of chitosan's maximum Ni adsorption capacities in the absence (1.00 mmol/g) and presence (1.49 mmol/g) of 500 mM $SO_4^{2-}$ and 10 mM Ca indicated the positive effect of these ions on Ni adsorption. The predominance of the neutrally charged $NiSO_4^0$ species in Ni/$SO_4^{2-}$ system has contributed to the enhanced Ni adsorption on chitosan as verified by X-ray photoelectron spectroscopy (XPS) analysis. However, kinetic studies confirmed that the Ni adsorption rate decreased by 4.5 times when $SO_4^{2-}$ was present. The subsequent continuous Ni adsorption from a real $SO_4^{2-}$-rich leachate in a column setup revealed that chitosan is selective for Ni over Ca and Cr with selectivity quotients of 9.6 ($K_{Ni/Ca}$) and 3.0 ($K_{Ni/Cr}$). Overall, this study indicated that Ni complexation with $SO_4^{2-}$ enhances the Ni adsorption capacity of chitosan, but slows down the adsorption process.

## 1. Introduction

Nickel (Ni) is one of the most versatile metals characterized by its workability and mechanical strength at high temperatures and in corrosive environments [1]. These properties make Ni indispensable in the manufacture of stainless steel, nonferrous alloys, rechargeable batteries, coins, paint formulation, and electroplating [2]. Being a mainstay of several economic sectors, primary Ni production has more than doubled from 1.14 Mt in 1999 to 2.70 Mt in 2019 [3, 4]. Substitution of Ni in several production processes has been reported to be feasible, although large-scale substitution would either be more expensive or result in a loss of product quality [3, 5]. Moreover, the economic geography of Ni mining and production has changed over time as it moved from Russia and Canada to New Caledonia, Australia, Brazil, and most recently to Indonesia and the Philippines [6]. Thus, sustainable nickel supply in the long run has become a global issue.

Considering that primary Ni resources continue to deplete and that new mining operations take at least a decade to plan and build, there is a need to recycle Ni from secondary resources, such as mine tailings and Ni-rich waste streams. Ni recycling does not only reduce the energy requirements and $CO_2$ emissions associated with primary Ni production, but also reduces the exposure to the unstable primary Ni market [7]. Furthermore, it can help prevent the buildup of Ni-rich waste streams that are mostly discharged to the environment, increasing the possibility of human and organism exposure to high levels of Ni [8].

Ni is present in many industrial effluents, including those of silver refineries, electroplating, zinc base casting, and storage battery industries [9], where its concentration usually ranges from 0.05 to 15.3 mM [10]. In many cases, these effluents are complex as there are other ions co-existing with Ni; two of the most ubiquitous being sulfate ($SO_4^{2-}$) and calcium (Ca) [11-14]. For instance, the majority of Ni plating industries utilizes sulfuric acid solutions to perform acid pickling for surface activation [11, 15]. Alternately, Ca is used in electroless Ni plating to precipitate the phosphite anion formed from the reaction of the hypophosphite reducing agent and $NiSO_4$ [16]. Ni, $SO_4^{2-}$, and Ca can also be co-existing in leachates, such as those derived from industrial sludge generated by $Ca(OH)_2$-neutralization of acidic wastewaters that are commonly produced by several industrial activities such as metal plating and mining [17]. Liquid streams associated with both Ni plating and industrial sludge leaching typically contain high amounts of $SO_4^{2-}$ and Ca that can reach up to 560 mM and 11 mM, respectively [14, 18]. Such high $SO_4^{2-}$ and Ca concentrations could present problems in some Ni removal and/or recovery technologies. In solvent extraction, for example, Ca could precipitate as gypsum, leading to scaling and blockages of the solvent extraction equipment [19]. Thus, alternative technologies that remain effective despite the complexity of the matrix should be developed for these $SO_4^{2-}$ and Ca-rich solutions.

Among the available technologies for Ni removal from wastewater, adsorption has shown great potential because of the following reasons: (1) easy handling, (2) cost efficiency due to possible adsorbent regeneration, (3) sustainability especially if the adsorbents used are abundant and low-cost by-products of other processes, and (4) potential selectivity for the target metal [20, 21]. Chitosan, for instance, has been used to adsorb metal ions from a wide range of effluent types, including wastewater from electroplating and metal-finishing operations, nickel-salt manufacturing plant, lead-battery manufacturing industry, and wool fabrics production [22, 23]. It is a polymeric material derived from the deacetylation of chitin, which is a component of the shells of various crustaceans [22, 24]. Due to the large volumes of shell wastes produced globally at around 6-8 Mt per year, their value-added applications should be promoted [25]. The ability of chitosan to bind metal ions is mainly due to its keto and both primary and secondary amine groups, which can serve as coordination sites for many metal ions [26]. Furthermore, the extent of metal adsorption depends on the source of chitosan, the degree of deacetylation, the nature of the metal ion, and the solution conditions such as pH [27]. Thus, metal adsorption on chitosan is highly unpredictable, making experimental procedures the only means to evaluate its adsorption capacity under particular conditions.

Ni adsorption on chitosan has been investigated in the past [28-31]; however, most of these studies did not report the pH change induced by chitosan, and its subsequent effect on Ni speciation. Thus, Ni adsorption on chitosan is usually overestimated and inaccurately interpreted. Furthermore, most Ni adsorption studies using chitosan focused on single metal solutions, and since other ions are usually co-existing with Ni at much higher concentrations, complex systems that are representative of real waste stream conditions should be investigated.

The main goal of this study was to test the hypothesis that the adsorption capacity of chitosan for Ni is largely affected by the presence of $SO_4^{2-}$, especially at pH values where Ni/$SO_4^{2-}$ complexes are formed. Batch adsorption tests were performed to adsorb Ni from a synthetic leachate containing Ni, $SO_4^{2-}$, and Ca. The effects of these ions on the Ni adsorption kinetics and maximum adsorption capacities of chitosan were studied in detail. Furthermore, the feasibility of using chitosan for selective Ni removal from a real leachate was assessed in a continuous flow system.

## 2. Materials and methods

### 2.1. Chitosan and chemicals

Chitosan with medium molecular weight and 75–85% degree of deacetylation was purchased from Merck, Darmstadt, Germany. Ni(II), Ca(II), $SO_4^{2-}$, and $NO_3^-$ stock solutions were prepared from Ni(NO$_3$)$_2$·6H$_2$O (UCB, Brussels, Belgium), Ca(NO$_3$)$_2$·4H$_2$O (Chem-Lab, Zedelgem, Belgium), Na$_2$SO$_4$ (Merck., Darmstadt, Germany), and NaNO$_3$ (Sigma-Aldrich, Missouri, USA) respectively. The working solutions used in the adsorption experiments were obtained by diluting the respective stock solutions. Deionized water (resistivity > 18.2 MΩ·cm) was used to prepare all aqueous solutions. If needed, the solution pH was adjusted using HNO$_3$.

### 2.2. Characterizations of chitosan

*Particle size distribution*

The particle size distribution of chitosan was determined by laser light scattering using a Malvern Mastersizer 2000 capable of measuring particle sizes ranging from 20 nm up to 2 mm. Chitosan was placed into the liquid dispersion unit until the signal was within the acceptable range of 15-25%. The test was run thrice and the average volume frequency (%) was plotted versus the particle size.

*Specific surface area*

The Brunauer-Emmett-Teller (BET) specific surface area (SSA) and the total pore volume (PV) of chitosan were determined through nitrogen adsorption-desorption measurements carried out at 77 K using a Micromeritics TriStar II 3020. The sample was degassed at 60°C under vacuum for 48 h prior to measurement.

*Isoelectric point (IEP)*

The IEP of chitosan was determined using SurPASS 3 electrokinetic analyzer equipped with a titration unit for automatic determination of the zeta potential as a function of pH. Approximately 50 mg of chitosan was placed in the powder cell of the equipment. The background electrolyte solution used was 0.01 M KCl, while 0.05 M HCl and 0.05 M KOH were used to adjust its pH.

*Fourier transform infrared spectroscopy*

The surface functional groups of chitosan before and after Ni(II) adsorption were examined using Fourier transform infrared spectroscopy (FTIR; Thermo Scientific Nicolet 6700 FT-IR Spectrometer). About 15 mg of chitosan was placed on top of potassium bromide (KBr) powder pressed in a micro-cup. The FTIR spectrum was collected within the wavenumber range of 700

– 4000 cm$^{-1}$ with spectral resolution of 4 cm$^{-1}$ and 256 scans. Pure KBr was scanned for background measurements, which were automatically subtracted from the sample spectrum.

*X-ray photoelectron spectroscopy*

X-ray photoelectron spectroscopy (XPS) analyses of raw and Ni-adsorbed chitosan in the absence and presence of SO$_4^{2-}$ were performed on a Kratos AXIS Supra X-ray photoelectron spectrometer using a monochromatic Al K(alpha) source (15mA, 15kV). The instrument work function was calibrated to give a binding energy (BE) of 83.96 eV for the Au 4f7/2 line for metallic gold and the spectrometer dispersion was adjusted to give a BE of 932.62 eV for the Cu 2p3/2 line of metallic copper. The Kratos charge neutralizer system was used on all samples. Survey scan analyses were carried out with an analysis area of 300 × 700 micrometers and a pass energy of 160 eV. High resolution analyses were carried out with an analysis area of 300 × 700 micrometers and a pass energy of 20 eV. Spectra have been charge corrected to the main line of the carbon 1s spectrum (C-C, C-H) set to 284.8 eV. Spectra were analyzed using CasaXPS software (version 2.3.14).

## 2.3. Chemical equilibrium modeling

Ni(II) speciation in aqueous systems with and without SO$_4^{2-}$ was modeled using Visual MINTEQ 3.0 [32]. The thermodynamic parameters of Visual MINTEQ were taken from the NIST Critical Stability constants database. The initial concentrations of Ni(II) and SO$_4^{2-}$ in each of the defined systems were used as the input to plot the fraction of the predominant Ni(II) species versus the solution pH. The obtained diagrams were used to observe any changes in Ni(II) speciation with increasing SO$_4^{2-}$ concentration.

## 2.4. Batch adsorption experiments

Batch experiments were used to investigate Ni(II) adsorption on chitosan under varying conditions. For the pH effect, 10 mM Ni(II) solutions with initial pH ranging from 4 to 8 were prepared. To achieve a liquid-to-solid (L/S) ratio of 100 mL/g, 10 mL of each solution was added to 100 mg of chitosan weighed in a polypropylene centrifuge tube. The initial pH of the suspensions was measured using a Thermo Scientific Orion Star A211 pH meter. Samples were transferred to an orbital shaker set at 115 rpm for 24 hours. Afterwards, the final pH of each sample was measured, followed by solid-liquid separation through filtration using membrane filters (0.45 µm pore size). The Ni(II) concentration in the filtrates was determined using inductively coupled plasma optical emission spectroscopy (Varian Vista-MPX CCD Simultaneous ICP-OES). All samples were prepared in quadruplicates and the data presented are mean values with standard deviation. Control samples without chitosan were also prepared and analyzed for Ni(II) concentration to monitor Ni(II) precipitation at different pH values.

Additionally, two sets of solutions containing—(1) 10 mM Ni(II) and 500 mM $SO_4^{2-}$, and (2) 10 mM Ni(II), 500 mM $SO_4^{2-}$, and 10 mM Ca(II)—were prepared to investigate the effects of $SO_4^{2-}$ and Ca(II) on Ni(II) adsorption at pH 7–8. All solutions also contained $NO_3^-$ and Na since Ni(II) and Ca(II) were added as nitrate salts, while $SO_4^{2-}$ was added as $Na_2SO_4$. The results of this experiment served as the basis of the following investigation examining the effect of increasing $SO_4^{2-}$ concentration on Ni(II) adsorption onto chitosan. For this adsorption experiment, five solutions containing 10 mM Ni(II) and varying $SO_4^{2-}$ concentrations (0.05 to 500 mM) were prepared at pH 7. The samples were prepared as previously described.

The removal efficiency and adsorption capacity of chitosan for Ni(II) were calculated using Eq.1 and 2, respectively.

$$R\ (\%) = \frac{C_0 - C_e}{C_0} \times 100 \tag{1}$$

$$q = \frac{(C_0 - C_e) \times V}{m} \tag{2}$$

Where $R$ is the Ni(II) removal efficiency (%), $q$ is the amount of Ni(II) adsorbed per unit mass of chitosan (mmol/g), $C_0$ and $C_e$ are the initial and equilibrium Ni(II) concentrations (mM), respectively, $V$ is the volume of the Ni(II) solution (L), and $m$ is the mass of chitosan (g).

Two batches of Ni(II) adsorption kinetic studies—with and without 500 mM $SO_4^{2-}$—were performed at pH 7 with an initial Ni(II) concentration of 10 mM. The samples were shaken for different time intervals ranging from 15 minutes to 24 hours. The pseudo-second order [33] kinetic model (Eq. 3) was considered to describe the time-dependent experimental data.

$$\frac{t}{q_t} = \frac{1}{k_2 q_e^2} + \frac{t}{q_e} \qquad (3)$$

Where $q_e$ and $q_t$ (mmol/g) are the Ni(II) adsorption capacities at equilibrium and time $t$, respectively, and $k_2$ (g/mmol·min) is the pseudo-second order rate constant of adsorption.

Ni(II) adsorption on chitosan was also assessed at different initial Ni(II) concentrations in systems with and without 500 mM $SO_4^{2-}$ and 10 mM Ca(II). The adsorption data were used to determine the maximum adsorption capacities ($q_{max}$) of chitosan for Ni(II) in both systems. The initial pH used was ca. 7.5, while the initial Ni(II) concentration was varied from 1 to 100 mM. The Langmuir adsorption isotherm model (Eq. 4) was fitted to the obtained experimental data.

$$q_e = \frac{q_{max} b C_e}{1 + b C_e} \qquad (4)$$

Where $q_e$ is the Ni(II) adsorption capacity at equilibrium (mmol/g), $q_{max}$ is the maximum adsorption capacity (mmol/g), $b$ is the Langmuir adsorption equilibrium constant (L/mmol), and $C_e$ is the equilibrium Ni(II) concentration (mM). The adsorption isotherms and the corresponding parameters were determined using SigmaPlot version 13.0.

### 2.5. Column adsorption experiment

The fixed bed adsorption experiment was carried out in a glass column with an inner diameter of 1 cm. The column was packed with 8.1 g chitosan of 10 cm bed height, and then fixed with glass beads on top. Prior to passing the test solution, the column was conditioned with deionized water for 30 minutes. The test solution was a real leachate [17] with components shown in Table 1.This was fed to the column at a constant upflow rate of 36 mL/h using a Watson-Marlow 530S peristaltic pump. Samples were taken at preset time intervals and their metal concentrations were analyzed by ICP-OES. The breakthrough curves were obtained by plotting the ratio of the metal concentration in the sample at any time *t* to the initial metal concentration in the leachate versus the time.

**Table 1.** Concentration (mM) of the compounds present in the real leachate (pH 7.6) used in the column experiment.

| Compound | Concentration (mM) |
| --- | --- |
| Ni | 1.10 |
| Ca | 11.7 |
| Co | 0.03 |
| Cr | 0.02 |
| Mn | 0.12 |
| $SO_4^{2-}$ | 563 |

The selectivity quotients ($K_{Ni/(Ca\ or\ Cr)}$), given by Eq. 5 [34], were calculated to assess the selectivity between Ni and the other ions (i.e., Ca or Cr) in the adsorption system.

$$K_{Ni/(Ca\ or\ Cr)} = \frac{K_d^{Ni}}{K_d^{Ca\ or\ Cr}} \qquad (5)$$

Where $K_d^{Ni}$ and $K_d^{Ca\ or\ Cr}$ are the distribution coefficients for Ni and Ca or Cr calculated using Eq. 6.

$$K_d = \frac{100 - X}{X} \cdot \frac{V}{m} \qquad (6)$$

Where *X* is the concentration of Ni, Ca, or Cr in the treated leachate expressed as a percentage of the initial concentration, *V* is the volume of the sampled leachate (mL), and *m* is the mass of chitosan (g).

3. **Results and Discussion**

**3.1. Chitosan characterization**

The physical and electrochemical characteristics of chitosan are summarized in Table 2. The volume median diameter $d_{50}$ of the chitosan used in this study is 0.54 mm, while 10% and 90% of the volume distribution ($d_{10}$ and $d_{90}$) have particle size diameters less than 0.17 and 1.15 mm, respectively. $N_2$ adsorption-desorption measurements confirmed the previous reports in the literature [22, 35] that chitosan has low specific surface area (0.8 m$^2$/g) and total pore volume (0.002 cm$^3$/g). The difference between the pH$_{PZC}$ and IEP of chitosan is positive (0.88), indicating a more negatively charged external than internal particle surface [36].

**Table 2.** Physical and electrochemical characteristics of chitosan.

| Properties | Chitosan |
|---|---|
| Particle size (mm) | $d_{10}$: 0.17<br>$d_{50}$  0.54<br>$d_{90}$: 1.15 |
| Specific surface area (m$^2$/g) | 0.8 |
| Total pore volume (cm$^3$/g) | 0.002 |
| pH$_{PZC}$ | 6.60[†] |
| IEP | 5.72 |

[†] Average of values reported in literature [37-40]

The FTIR spectrum of chitosan displayed different characteristic peaks suggesting which functional groups are present on the chitosan surface (Figure 1). The broad band at around 3330 cm$^{-1}$ suggests overlapping and vibrational stretching of the N–H bond of the secondary amide and O–H bonds of the primary alcohol groups. The peaks at 2878 cm$^{-1}$, 1657 cm$^{-1}$, and 1596 cm$^{-1}$ could be attributed to C–H stretching, C=O stretching of the amide, and N–H bending of non-acetylated amides, respectively. Other remarkable peaks at 1422 cm$^{-1}$ and 1382 cm$^{-1}$ correspond to C–O–C stretching vibrations and CH$_3$ symmetrical deformations, respectively. The band at around 1013 cm$^{-1}$ could be assigned to the vibrational stretching of the C–O bonds of carboxylic acid groups [31, 37, 41]. Chitosan is well known for the amine subgroups of its amides, which primarily influence the adsorption of transition metals by coordination [22]. The FTIR spectra of chitosan after Ni(II) and Ni(II)/SO$_4^{2-}$ adsorption (Figure 1) indicated a shift in some of the characteristic peaks, specifically those at 3330 cm$^{-1}$, 1422 cm$^{-1}$, and 1382 cm$^{-1}$. These results imply the involvement of amide, alcohol, ether, and alkane functional groups of chitosan in the adsorption of Ni(II).

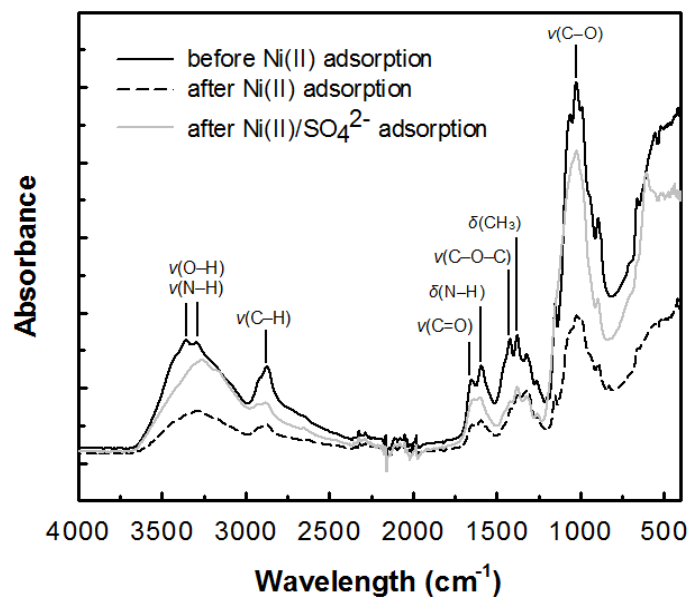

**Figure 1.** FTIR spectra of chitosan before Ni(II) adsorption, after Ni(II) adsorption, and after Ni(II)/SO$_4^{2-}$ adsorption showing peak assignments from 4000-400 cm$^{-1}$, where $v$ and $\delta$ correspond to the stretching and bending vibrations, respectively.

### 3.2. Effect of pH on Ni(II) removal

The solution pH is a key parameter in metal adsorption processes as it affects both metal speciation and the adsorbent's surface characteristics [42]. In the solution containing Ni(II) with no additional SO$_4^{2-}$ or Ca(II), the Ni(II) removal efficiency of chitosan slightly decreased from 42% to 31% as the final pH increased from 5.6 to 6.7, and then increased again up to 44% as the final pH increased to 7.82 (Figure 2). The small decrease in Ni(II) removal (11%) between pH 5.6 and 6.7 was unexpected considering that chitosan has an IEP of 5.72. At pH values higher than the IEP, the chitosan surface becomes negatively charged, which should favor a stronger electrostatic interaction with Ni$^{2+}$, the predicted predominant Ni(II) species at pH ≤ 8 (Figure 3A). This result contrasts with previous literature wherein Ni(II) adsorption on chitosan increases with pH due to the decreasing competition for adsorption surface sites between $H^+$ and $Ni^{2+}$ [43, 44]. In the present study, the higher Ni(II) adsorption at lower pH values could be due to chitosan solubilization at acidic pH, resulting in a better accessibility for interaction of metals with the keto and amine group of the 2-amino-2-deoxy-D-glucose (glucosamine) unit [45]. The increase in Ni(II) removal beyond pH 6.7 could be attributed to the electrostatic attraction between $Ni^{2+}$ and the negatively charged chitosan surface due to the deprotonation of some of its functional groups. For instance, the amine groups of chitosan deprotonate at pH above their $pK_a$ of 6.2-6.5 [46]. Furthermore, the loss of hydrogen ions in chitosan at weakly alkaline solutions may result in the coordination of $Ni^{2+}$ ions to hydroxyl oxygen and nitrogen atoms [47]. In addition, it is highly possible that the Ni(II) removal at pH 7.82 was partly due to precipitation. The chitosan-free control samples showed a Ni(II) removal of 7% at a final pH of 7.48 (Table SI-1). Likewise, the chemical speciation of Ni(II) in aqueous solution (Figure 3A) also shows that solid $Ni(OH)_2$ starts to form at around pH 7.5.

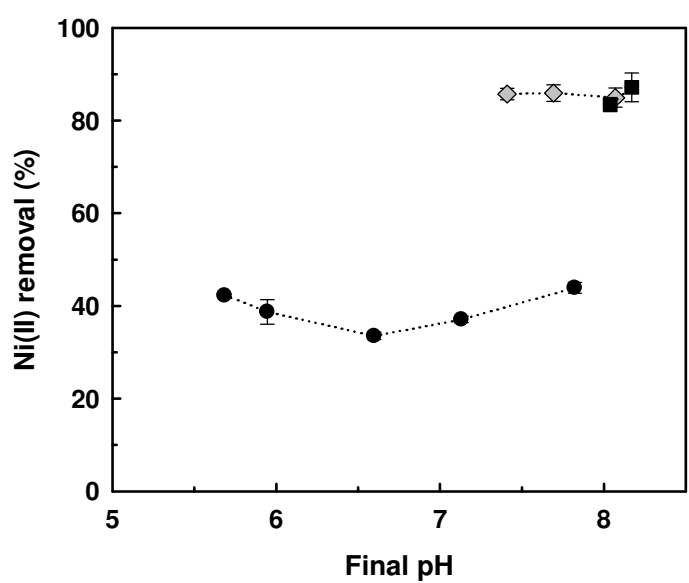

**Figure 2.** Effect of pH on Ni(II) removal using chitosan in a system (●) without $SO_4^{2-}$ and Ca(II), (♦) with 500 mM $SO_4^{2-}$, and (■) with 500 mM $SO_4^{2-}$ and 10 mM Ca(II). Ni(II) removal

is shown as the average value of the replicates (number of samples, N = 4) with error bars representing the standard deviation. Where not shown, error bars are within the size of the symbol. Experimental conditions: 10 mM initial Ni(II) concentration, 100 mL/g L/S ratio, room temperature, 24 h contact time. Note: The lines were added for visual clarity only.

The effect of $SO_4^{2-}$ and Ca(II) on Ni(II) adsorption onto chitosan was aimed to be tested at the same pH range as that of the system with Ni(II) in the absence of $SO_4^{2-}$ and Ca(II) (Figure 2). However, the addition of $SO_4^{2-}$ and Ca(II) resulted in higher final pH values, making it difficult to objectively compare the Ni(II) removal efficiencies among the three different systems with Ni(II); Ni(II) and $SO_4^{2-}$; and Ni(II), $SO_4^{2-}$, and Ca(II). The presence of 500 mM $SO_4^{2-}$ increased the Ni(II) removal efficiency of chitosan by at least 43%, which is reasonably high considering the presence of 1000 mM Na in solution since $SO_4^{2-}$ was added as $Na_2SO_4$. The presence of excess $Na^+$ could compete for the surface adsorption sites and decrease metal adsorption [48]. In the presence of 500 mM $SO_4^{2-}$, $Ni(OH)_2$ precipitation only starts at around pH 8 as estimated from the chemical speciation modeling (Figure 3B). The control samples with 10 mM Ni(II) and 500 mM $SO_4^{2-}$ (Table SI-1) have also shown that there was no precipitation up to a final pH of 8.24. These suggest that the reported removal efficiencies of chitosan can be solely attributed to adsorption without any contribution from precipitation. Further addition of 10 mM Ca(II) to the system with 500 mM $SO_4^{2-}$ and 10 mM Ni(II) did not present an apparent effect on Ni(II) removal, although it is possible that there was $Ni(OH)_2$ precipitation at around pH 8.

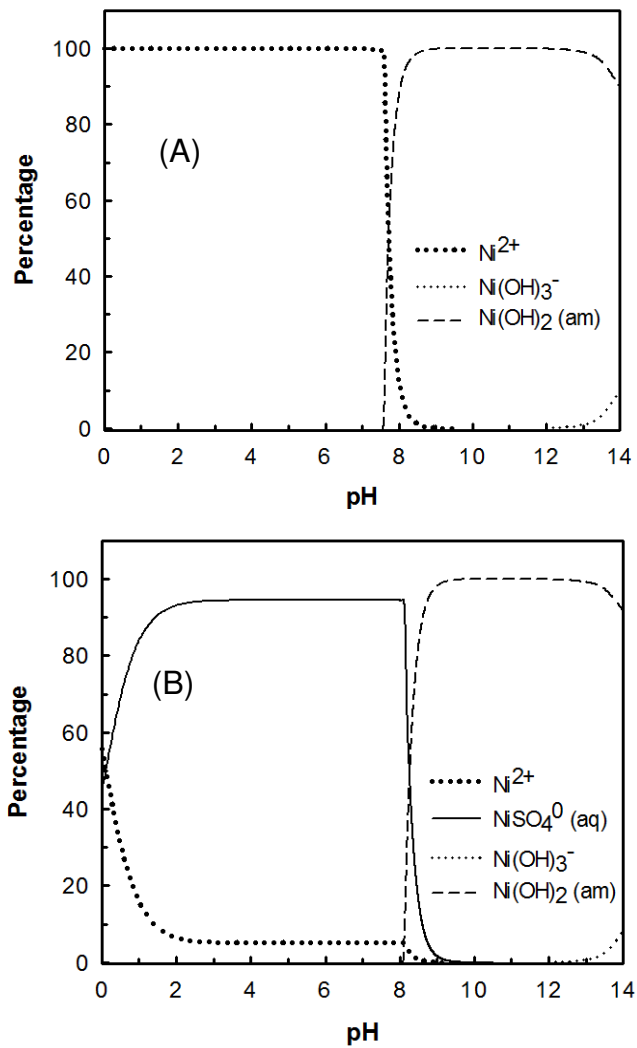

**Figure 3.** Equilibrium chemical speciation of 10 mM Ni(II) in aqueous solution (A) without $SO_4^{2-}$ and (B) with 500 mM $SO_4^{2-}$ estimated using Visual MINTEQ.

## 3.3. Effect of $SO_4^{2-}$ on Ni(II) adsorption

As it was evident that $SO_4^{2-}$ increases Ni(II) adsorption onto chitosan, the extent of this enhancement was further investigated. Figure 4 shows the Ni(II) removal efficiencies of chitosan from solutions containing various $SO_4^{2-}$ concentrations. Lower $SO_4^{2-}$ concentrations (i.e., 0.05-0.5 mM) did not have an effect on Ni(II) adsorption onto chitosan, but as $SO_4^{2-}$ concentration increased to 5 mM, Ni(II) removal increased from 32% to 67%. Higher Ni(II) removal (89%) was further achieved when 50 mM $SO_4^{2-}$ was present in the system, but beyond this level of $SO_4^{2-}$, Ni(II) removal did not further increase. It was also observed that the final solution pH increased with increasing $SO_4^{2-}$ concentration (Figure 4). The solution pH increased upon contact with chitosan, for which the increase also depends on the activity of the components present in the solution. With $SO_4^{2-}$ present, it is expected that the activity of $Ni^{2+}$ as a Lewis acid will be lower as it coordinates with $SO_4^{2-}$ instead of water molecules; hence, the increasing trend in pH. Nevertheless, the control samples did not show any Ni(II) removal even in the solution with the highest $SO_4^{2-}$ concentration.

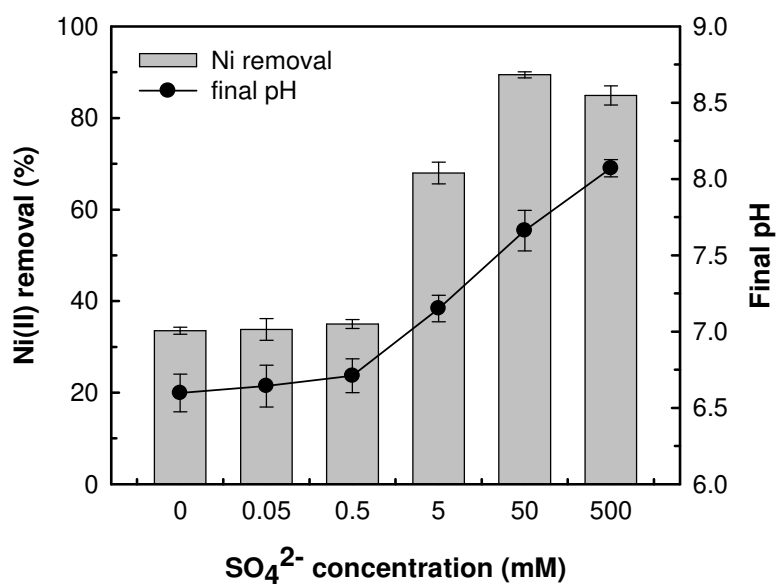

**Figure 4.** Effect of increasing $SO_4^{2-}$ concentration on Ni(II) removal efficiency of chitosan. Ni(II) removal or final pH is shown as the average value of the replicates (number of samples, N = 4) with error bars representing the standard deviation. Experimental conditions: 10 mM initial Ni(II) concentration, 100 mL/g L/S ratio, initial pH 7, room temperature, 24 h contact time.

Mitani, et al. [29] also observed an increasing trend in the Ni(II) removal efficiency of swollen chitosan beads as $SO_4^{2-}$ concentration in the system increased. The Ni(II) speciation diagrams obtained from the experimental conditions also revealed an increasing fraction of $NiSO_4^0$ species as $SO_4^{2-}$ concentration increased, which is similar to what was observed in our study (Figure SI-1). Furthermore, Mitani, et al. [29] suggested that $SO_4^{2-}$ may have caused some conformational change of chitosan, making it easy for the Ni(II) ions to interact with the chitosan surface. $SO_4^{2-}$ ions tend to accumulate in the chitosan chains and crosslink the protonated amine groups, introducing negative charges that are capable of complexing metal cations [49].

### 3.4. X-ray photoelectron spectroscopy analysis

XPS analysis of chitosan tried to reveal the differences in the chemical states of elements before and after Ni(II) adsorption and verify the effect of $SO_4^{2-}$ present. The resulting spectra in the binding energy range of 0-1200 eV are plotted in Figure 5. The XPS spectrum of chitosan before adsorption clearly demonstrates the existence of oxygen (O 1s), nitrogen (N 1s), calcium (Ca 2p), and carbon (C 1s) on the surface of chitosan. The N 1s peak was resolved into two peaks at 399.3 and 401.5 eV (Figure SI-2A), indicative of amine/amide groups and graphitic N, respectively. The details of different carbon functionalities on chitosan surface were also determined from the C 1s high-resolution scan (Figure SI-2B), which was deconvoluted into four component peaks, namely O–C=O (289.1 eV), C=O (287.8 eV), C–OH (286.3 eV), and C–C, C–H (284.8 eV). These N and C functionalities corresponded to the obtained FTIR results in Figure 1.

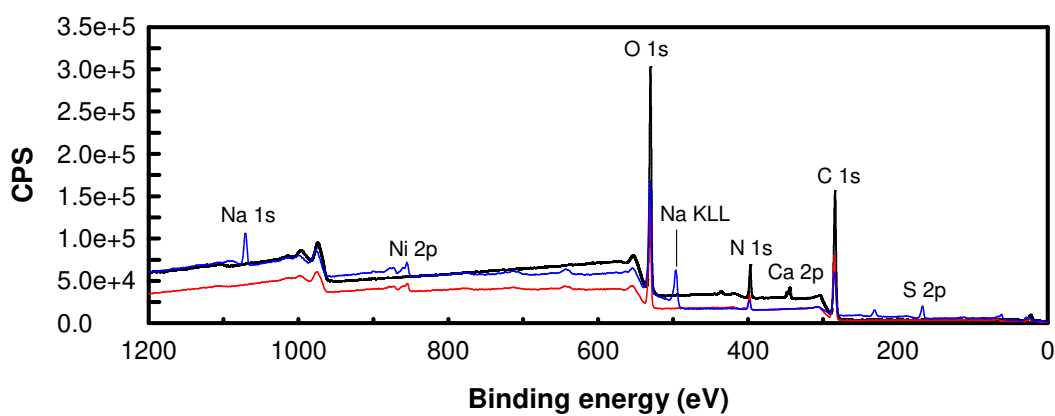

**Figure 5.** XPS survey spectra of chitosan before (black) and after 10 mM Ni(II) adsorption in the absence (red) and presence (blue) of 500 mM $SO_4^{2-}$. The spectra are plotted as counts per second (CPS) versus the binding energy.

In addition to C, O, and N, Ni was also observed in the XPS spectra of chitosan after Ni(II) adsorption (Figure 5). The high-resolution scans and deconvolution of the Ni 2p peaks observed in the absence and presence of $SO_4^{2-}$ are shown, respectively, in Figures 6A and 6B, which both display a main photo-peak and an associated satellite peak located at 5.8 eV higher than the main peak. In the absence of $SO_4^{2-}$, the peaks at 856.0 eV and 861.8 eV of the main and satellite peaks, respectively, are attributed to Ni(OH)$_2$, while those at 855.3 eV and 860.8 eV are attributed to NiO. With $SO_4^{2-}$ present, the Ni 2p peak was resolved into major peaks indicative of Ni(OH)$_2$ (855.9 eV, 861.7 eV) and NiSO$_4$ (856.5 eV, 861.5 eV). Figure SI-3 illustrates that the binding energy of N 1s (amine groups) increased from 399.3 eV to 399.8 eV and 399.6 eV after Ni(II) adsorption in the absence and presence of $SO_4^{2-}$, respectively. Similarly, the binding energy of the major C 1s peak corresponding to C–OH increased from 286.3 eV to 286.5 eV (without $SO_4^{2-}$) and 286.4 eV (with $SO_4^{2-}$) (Figure SI-4).

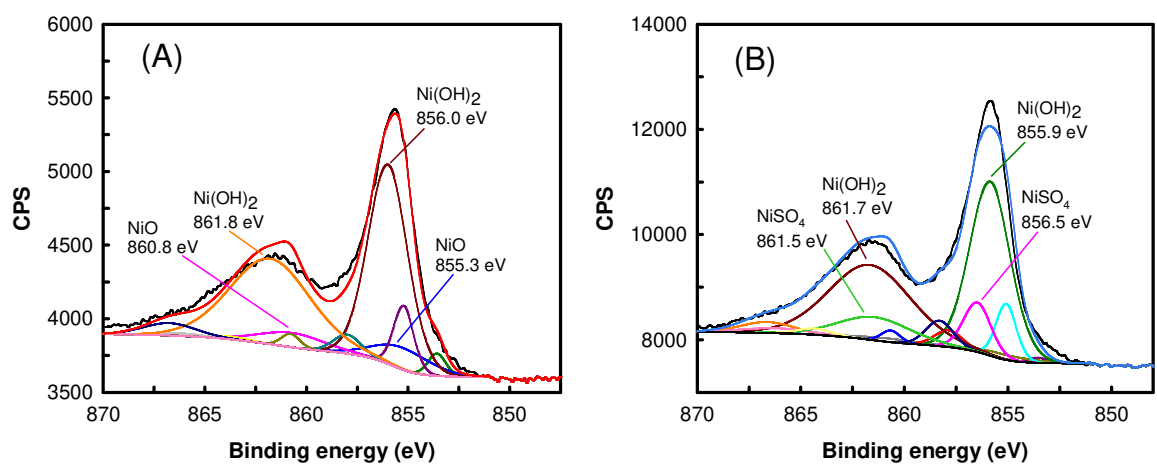

**Figure 6.** High-resolution XPS scan spectra over Ni 2p of chitosan after Ni(II) adsorption in the (A) absence and (B) presence of $SO_4^{2-}$. The spectra are plotted as counts per second (CPS) versus the binding energy.

The XPS results indicate that Ni(II) speciation, specifically the formation of Ni(SO$_4$)$^0$, promoted Ni(II) adsorption onto chitosan. Furthermore, Ni(II) and $SO_4^{2-}$ complexation was experimentally validated when non-complexing nitrate was used instead of $SO_4^{2-}$ and did not result in increased Ni(II) removal (Figure SI-5). In Figure 6B, the satellite peak for Ni and the measured binding energy (856.5 eV) for $NiSO_4$ in the main peak are both characteristics of a paramagnetic Ni(II) in an octahedral environment [50]. This can be supported by the findings of Matienzo, et al. [51] that a binding energy within 856-857 eV is characteristic of octahedral Ni(II) and that $NiSO_4 \cdot 6H_2O$ has a binding energy equivalent to 856.5 eV. Thus, the obtained XPS data suggest that the adsorbed $NiSO_4$ was an aquated species, specifically a hexahydrate ion (Figure SI-6). It is then likely that this Ni(II) species was adsorbed through $SO_4^{2-}$ binding with the amine groups of chitosan, which can be supported by the FTIR (Figure 1) and XPS (Figures SI-2 and 3) results.

The presence of $Ni(OH)_2$ on both chitosan samples as suggested by the XPS results does not correspond to the control samples and chemical speciation modeling results. Additional speciation modeling using PHREEQC (Figure SI-7) also showed that $Ni(OH)_2$ starts to form only at pH 7.6 and pH 8.1 in the absence and presence of $SO_4^{2-}$, respectively. The mismatch between the XPS results and those from the control samples and chemical speciation modeling suggests that $Ni(OH)_2$ is only formed when chitosan is present. It is possible that $Ni(OH)_2$ was a surface-induced precipitate because of the micro-conditions on the surface of chitosan, such as local deviations in pH.

### 3.5. Ni(II) adsorption kinetics

The effect of $SO_4^{2-}$ on the Ni(II) adsorption rate of chitosan was investigated at different time intervals. Figure 7 shows the ratio of the Ni(II) adsorption capacity at time $t$ ($q_t$) to the equilibrium Ni(II) adsorption capacity ($q_e$) of chitosan as a function of time. In the first 15 minutes of adsorption, chitosan achieved a higher $q_t/q_e$ for the system without $SO_4^{2-}$ (0.4) compared to the system with 500 mM $SO_4^{2-}$ (0.2). After 15 minutes up to around 4 h, both systems achieved almost similar $q_t/q_e$, suggesting that the adsorption rates are comparable. Equilibrium was reached after 4 and 16 h in the absence and presence of 500 mM $SO_4^{2-}$, respectively. The rapid Ni(II) removal during the first hour of adsorption (Figure 7) could be partly attributed to the rapid swelling of chitosan, which provided more internal surface for adsorption.

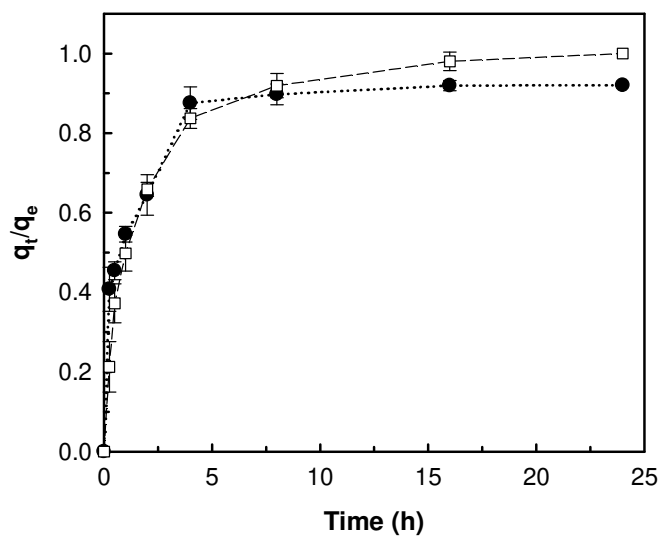

**Figure 7.** Kinetics of Ni(II) adsorption on chitosan in systems without $SO_4^{2-}$ (●, dotted line) and with 500 mM $SO_4^{2-}$ (□, dashed line). The ratio of the adsorption capacity at time $t$ ($q_t$) to the equilibrium adsorption capacity ($q_e$) of chitosan for Ni(II) ($q_t/q_e$) is shown as the average value of the replicates (number of samples, N = 4) with error bars representing the standard deviation. Where not shown, error bars are within the size of the symbol. The lines were added for visual clarity only. Experimental conditions: 10 mM initial Ni(II) concentration, 100 mL/g L/S ratio, initial pH 7, room temperature.

The adsorption kinetics data of chitosan were fitted to the pseudo-second order kinetic model (Figure SI-8). This model obtained good estimates of the experimental $q_e$ values, which are 0.43 mmol/g and 0.87 mmol/g for the systems without and with 500 mM $SO_4^{2-}$, respectively (Table 3). This implies that the rate limiting step may be chemical adsorption, suggesting monolayer adsorption of $NiSO_4^0$ species onto chitosan [52]. This is in contrast to what was suggested by Mende, et al. [28] that multilayer Ni(II) adsorption occurs in the presence of $SO_4^{2-}$. The adsorption rate constant $k_2$ confirmed that Ni(II) adsorbed more quickly onto chitosan when $SO_4^{2-}$ was not present, although the amount of Ni(II) adsorbed in the former system was less.

**Table 3.** Kinetics parameters for Ni(II) adsorption onto chitosan.

| Pseudo-second order model parameters | Without $SO_4^{2-}$ | With 500 mM $SO_4^{2-}$ |
|---|---|---|
| $q_e$ (mmol/g) | 0.379 | 0.888 |
| $k_2$ (g/(mmol·min)) | 0.080 | 0.018 |
| $R^2$ | 0.986 | 0.997 |

## 3.6. Adsorption isotherms

The Langmuir isotherm model was fitted to the equilibrium data of Ni(II) adsorption onto chitosan in the absence and presence of both $SO_4^{2-}$ and Ca(II) (Figure 8). The isotherms for both systems displayed different curve progressions. Without the coexisting ions $SO_4^{2-}$ and Ca(II), the Langmuir isotherm featured a minor curvature that is ranging over the entire isotherm. On the contrary, the isotherm for the system containing $SO_4^{2-}$ and Ca(II) displayed a steep increase at lower Ni(II) concentrations, followed by a rapid saturation. In general, the Langmuir model is more suitable for isotherms like those of the system with $SO_4^{2-}$ and Ca(II) as this model indicates a monolayer coverage and energetic homogeneity of the adsorption sites [28]. Table 4 shows the parameters from the fittings of the Langmuir model, which showed a good fit of the adsorption data for both systems. The estimated maximum Ni(II) adsorption capacities ($q_{max}$) of chitosan revealed a higher $q_{max}$ (1.49 mmol/g) when $SO_4^{2-}$ and Ca(II) are present in the solution (Table 4). A larger Langmuir adsorption equilibrium constant (b) of 1.50 L/mmol was also obtained in the presence of $SO_4^{2-}$ and Ca(II) compared to 0.08 L/mmol when only Ni(II) was present. From these equilibrium constants, negative values of standard Gibbs free energy ($\Delta G°$, Eq. SI-1) of -17.94 kJ/mol and -10.75 kJ/mol were obtained for the respective systems, indicating that the adsorption process was spontaneous. The larger $\Delta G°$ in the presence of $SO_4^{2-}$ and Ca(II) suggests that when greater quantities of reactants are employed in the reaction, the released Gibbs free energy in the system will also increase, indicating a higher adsorption driving force [53, 54].

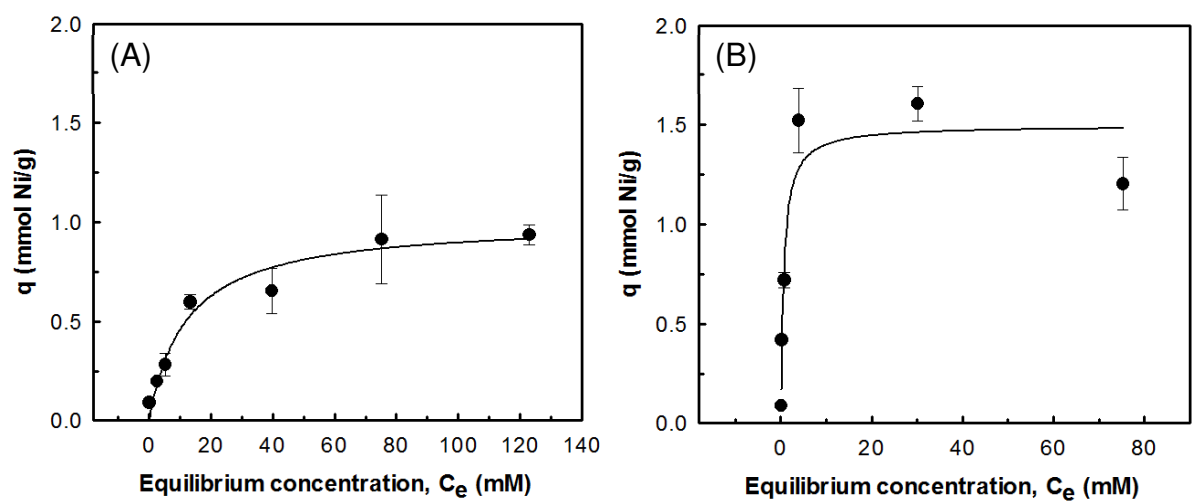

**Figure 8.** Langmuir isotherm model (solid line) of Ni(II) adsorption onto chitosan in a system (A) without $SO_4^{2-}$ and Ca(II) and (B) with 500 mM $SO_4^{2-}$ and 10 mM Ca(II). The amount of Ni(II) adsorbed (mmol) per gram of chitosan (q) is shown as the average value of the replicates (number of samples, N = 4) with error bars representing the standard deviation. Where not shown, error bars are within the size of the symbol. The large error bars at some concentrations were due to the large dilution factor applied to prepare the samples for analysis.

**Table 4.** Constants and correlation coefficients of the Langmuir model for Ni(II) adsorption using chitosan.

| Langmuir model parameters | Without $SO_4^{2-}$ and Ca(II) | With 500 mM $SO_4^{2-}$ and 10 mM Ca(II) |
|---|---|---|
| $q_{max}$ (mmol/g) | 1.00 | 1.49 |
| $b$ (L/mmol) | 0.08 | 1.50 |
| $R^2$ | 0.96 | 0.90 |

The obtained $q_{max}$ values of chitosan are comparable with those found in the literature concerning Ni(II) adsorption (Table 5). In some cases, chitosan exhibited higher $q_{max}$ compared to the other adsorbents used in Ni(II) adsorption without $SO_4^{2-}$ in the system, such as *Saccharomyces cerevisiae* and chitosan(chitin)/cellulose composites, which both achieved a $q_{max}$ of 0.2 mmol/g [55, 56]. In addition, all studies shown in Table 5 did not report the final equilibrium pH, which could provide information if the reported $q_{max}$ was overestimated due to $Ni(OH)_2$ precipitation. Thus, the reported $q_{max}$ values could possibly be lower compared to the actual $q_{max}$ caused by adsorption alone. Other adsorbents exhibited higher adsorption capacities than chitosan, specifically the polymeric resins Amberlite IRC 748® and Dowex M4195®, which were also tested in solutions containing $SO_4^{2-}$ [57].

**Table 5.** Comparative maximum Ni(II) adsorption capacities of different adsorbents.

| Adsorbent | Condition | pH | $q_{max}$ (mmol/g) | Reference |
|---|---|---|---|---|
| *Chlorella vulgaris* | Only Ni(II) in solution | 4.5[a] | 1.0 | Aksu [58] |
| *Sargassum muticum* | Only Ni(II) in solution | 5.0[a] | 1.2 | Bermúdez, et al. [59] |
| *Gracilaria caudata* | Only Ni(II) in solution | 5.0[a] | 0.8 | Bermúdez, et al. [59] |
| *Oedogonium hatei* (treated with 0.1 M HCl) | Only Ni(II) in solution | 5.0[a] | 0.8 | Gupta, et al. [60] |
| Activated carbon from waste apricot | Only Ni(II) in solution | 5.0[a] | 1.7 | Erdoğan, et al. [10] |
| Activated carbon from coirpith | Only Ni(II) in solution | 5.0[a] | 1.1 | Kadirvelu, et al. [9] |
| *Saccharomyces cerevisiae* | Only Ni(II) in solution | 6.8[a] | 0.2 | Padmavathy, et al. [55] |
| Amberlite IRC 748® | With $SO_4^{2-}$ * | 3.0[a] | 2.1 | Mendes and Martins [57] |
| Dowex M4195® | With $SO_4^{2-}$ * | 3.0[a] | 1.6 | Mendes and Martins [57] |
| Ionac SR-5® | With $SO_4^{2-}$ * | 3.0[a] | 1.4 | Mendes and Martins [57] |

| Chitosan/magnetite nanocomposite beads | Only Ni(II) in solution | 6.0[a] | 0.9 | Tran, et al. [31] |
| Chitosan(chitin)/cellulose composites | Only Ni(II) in solution | 5.3[a] | 0.2 | Sun, et al. [56] |
| Chitosan | Only Ni(II) in solution | 6.8-7.3[b] | 0.9 | This study |
| Chitosan | With 500 mM $SO_4^{2-}$ and 10 mM Ca(II) | 7.1-8.4[b] | 1.4 | This study |

[a] Initial pH value. Authors did not report final equilibrium pH.
[b] Final equilibrium pH. Authors observed a decreasing pH trend with increasing Ni(II) concentration in solution.
\* $SO_4^{2-}$ concentration was not reported.

### 3.7. Column experiment

Since chitosan has shown better Ni(II) adsorption in the presence of $SO_4^{2-}$ and Ca(II), a column test was carried out to verify its performance in a semi-continuous system (Figure 9). Because it was practically not possible to keep the column test running overnight, the flow was stopped as indicated by the gray areas in Figure 9. The composition of the leachate solution used in the column test is given in Table 1. The leachate has a pH of 7.6, which then varied from 7.4 to 8.0 after adsorption. $SO_4^{2-}$ (563 mM) and Ca(II) (11.7 mM) are present in much higher concentrations compared to Ni(II) (1.1 mM), which is a good representation of real secondary resources conditions. Despite the presence of high Ca(II) and $SO_4^{2-}$ concentrations, the results show that Ni(II) can be effectively removed at 95.9-100% in the first 7 h, corresponding to a percolate-to-inlet Ni(II) concentration ratio ($C_t/C_0$) of < 0.05 (Figure 9). The decrease in Ni(II) $C_t/C_0$ observed at the start of sampling days 2 and 3 was probably due to longer contact time between chitosan and the sampled leachate when the flow was stopped. Chitosan was also found to be selective for Ni(II) over Ca(II) and Cr(III). This selective Ni(II) adsorption was also reflected by the calculated selectivity quotients $K_{Ni/Ca}$ (9.6) and $K_{Ni/Cr}$ (3.0) for the last time point, which were even higher in the initial phase of the column experiment. However, due to much higher Ca(II) concentration in the influent compared to Ni(II) concentration, Ca(II) was still adsorbed by the chitosan in the column. This needs further investigation if the goal is to obtain pure Ni(II) after the adsorption process.

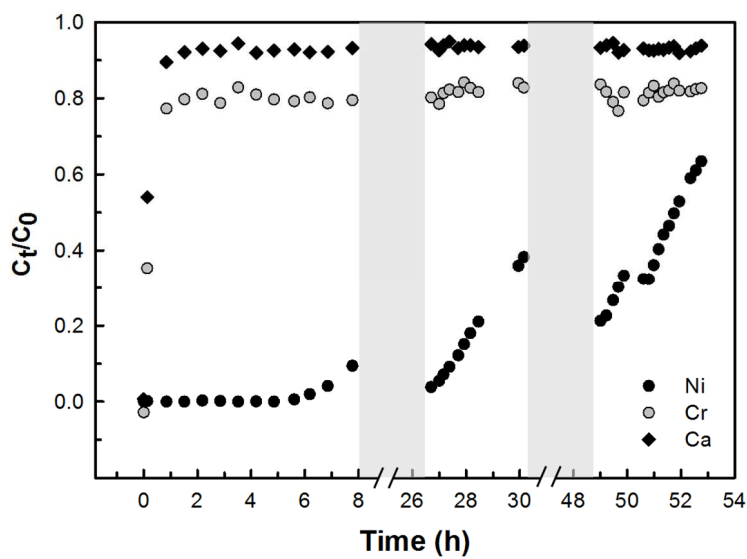

**Figure 9.** Breakthrough curves for Ni(II), Cr(III), and Ca(II) adsorption onto chitosan. $C_t$ and $C_0$ correspond to the percolate metal concentration at any time $t$ and the inlet metal concentration, respectively.

At the last collected time point (52.7 h), the column still adsorbed 36.6% of the inlet Ni, indicating that saturation was still not reached. At this time point, Ni(II) loading onto chitosan was only 0.04 mmol/g. The saturation of chitosan was not reached because of the low contact time between the leachate and chitosan in the column. For the same amount of leachate treated, the amount of Ni(II) adsorbed could be increased by reducing the flow rate of the leachate for longer contact time, applying sequential adsorption through columns in series, or increasing the size of the column. In terms of sustainability, further investigations are needed to recover both Ni and chitosan. Some studies have successfully addressed metal desorption from chitosan, however, this needs to be carefully performed because of the limiting low pH conditions that can be used to avoid solubility of chitosan [22]. Additionally, multicomponent modeling could be useful in designing integrated Ni(II) removal and chitosan regeneration systems to assess their feasibility for future industrial applications where complex matrices are usually found.

## 4. Conclusions

This study has investigated the potential of chitosan in removing Ni(II) from $SO_4^{2-}$ and Ca(II)-rich solutions that are representative of real secondary resources conditions. The analysis of Ni(II) adsorption equilibrium data in the absence and presence of $SO_4^{2-}$ and Ca(II) could be best described by the Langmuir adsorption isotherm, suggesting that the Ni(II) species were adsorbed in monolayer onto the adsorption sites of chitosan. Results showed that the chitosan obtained a higher maximum Ni(II) adsorption capacity of 1.49 mmol/g in the presence of 500 mM $SO_4^{2-}$ and 10 mM Ca(II) compared to 1.00 mmol/g when these ions are not present. XPS results indicated that the enhanced Ni(II) adsorption in the presence of $SO_4^{2-}$ was due to the formation of $NiSO_4^0$ in solution. Although less Ni(II) was adsorbed in the absence of $SO_4^{2-}$, the kinetics data showed that the Ni(II) adsorption rate was faster in this system, achieving equilibrium after 4 h. Furthermore, chitosan has shown selectivity for Ni(II) over Ca(II) and Cr(III) as validated in a continuous column setup using a real leachate. This is highly favorable especially if pure Ni(II) can be obtained as it can reduce the operational costs associated with further downstream processing. Overall, this study demonstrated the potential of chitosan for Ni(II) adsorption from $SO_4^{2-}$- and Ca(II)-rich streams without the pre-adsorption treatments possibly needed due to the complexity of these streams. For future investigations, the column setup and its operating conditions should be optimized to achieve full Ni(II) selectivity, which is essential for Ni(II) recovery.

**Acknowledgements**

The authors would like to acknowledge Liesbeth Horckmans (Vlaamse Instelling voor Technologisch Onderzoek, VITO) for providing the leachate used in this study, Prof. Pascal

Van Der Voort (Centre for Ordered Materials, Organometallics and Catalysis, UGent) for allowing us to use their FTIR Spectrometer, and Prof. Loretta Li (Environmental Engineering Group, UBC) for allowing us to use their laboratory for the additional experiments needed for the revision.

**Funding sources**

This research was supported by the EU Horizon 2020 METGROW+ project (Grant Agreement n° 690088) on Metal Recovery from Low Grade Ores and Wastes. NRN is funded by UGent Special Research Fund (BOF). KF is funded by SBO Project SMART (Sustainable Metal Extraction from Tailings) that fits in the SIM (Strategic Initiative Materials) program of Flanders.**References**

[1] A. Nieto, The Strategic Importance of Nickel: Scenarios and Perspectives Aimed to Global Supply, Journal of Mining and Metallurgy Section B Metallurgy, 332 (2013) 510-518.

[2] V. Srivastava, C.H. Weng, V.K. Singh, Y.C. Sharma, Adsorption of Nickel Ions from Aqueous Solutions by Nano Alumina: Kinetic, Mass Transfer, and Equilibrium Studies, Journal of Chemical & Engineering Data, 56 (2011) 1414-1422.

[3] USGS, Mineral commodity summaries 2020, in: Mineral Commodity Summaries, Reston, VA, 2020, pp. 204.

[4] USGS, Mineral commodity summaries 2000, in: Mineral Commodity Summaries, Reston, VA, 2000.

[5] M.L.C.M. Henckens, E. Worrell, Reviewing the availability of copper and nickel for future generations. The balance between production growth, sustainability and recycling rates, Journal of Cleaner Production, 264 (2020) 121460.

[6] X. Zeng, M. Xu, J. Li, Examining the sustainability of China's nickel supply: 1950–2050, Resources, Conservation and Recycling, 139 (2018) 188-193.28

# Supplementary Information

# Selective and enhanced nickel adsorption from sulfate- and calcium-rich solutions using chitosan

Nina Ricci Nicomel[a,*], Lila Otero-Gonzalez[a,I], Karel Folens[b], Bernd Mees[a], Tom Hennebel[b], Gijs Du Laing[a]

[a] Laboratory of Analytical Chemistry and Applied Ecochemistry, Department of Green Chemistry and Technology, Ghent University, Coupure Links 653, 9000 Ghent, Belgium

[b] Center for Microbial Ecology and Technology (CMET), Department of Biotechnology, Ghent University, Coupure Links 653, 9000 Ghent, Belgium

* Corresponding author:

E-mail: NinaRicci.Nicomel@UGent.be

Tel.: +3292646131

Fax: +3292646232

Present address:

[I] IDENER, Earle Ovington 24, 8-9, 41300 La Rinconada, Seville, Spain

**Standard Gibbs free energy**

The standard Gibbs free energy (ΔG°) was calculated using the following equation:

$$\Delta G^0 = -R\,T\,\ln(b) \qquad (Eq.\,SI-1)$$

where $R$ is the universal gas constant (8.314 J mol$^{-1}$ K$^{-1}$), $T$ is the temperature (K) and $b$ is the adsorption equilibrium constant obtained from the Langmuir isotherm model.

**Table SI-1.** Measured Ni(II) removal (%) in the control samples at different experimental conditions.

| Control sample | Ni(II) (mM) | $SO_4^{2-}$ (mM) | Ca(II) (mM) | Average initial pH | Average final pH | Average Ni(II) removal (%) |
|---|---|---|---|---|---|---|
| 1 | 10 | 0 | 0 | 7.48 | 7.48 | 6.99 |
| 2 | 10 | 500 | 0 | 8.06 | 8.24 | 0 |
| 3 | 10 | 500 | 10 | 7.87 | 7.80 | 0 |

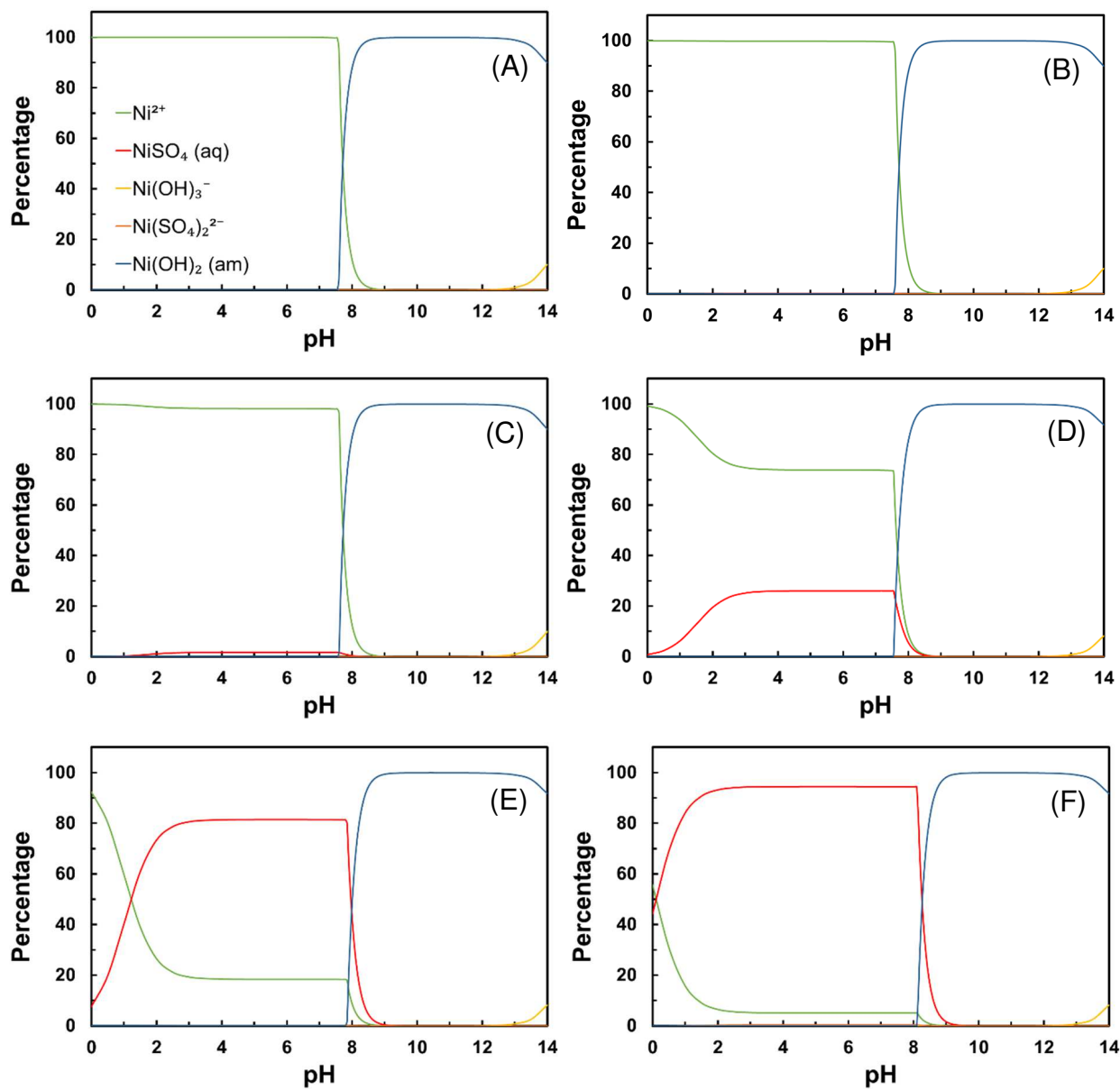

**Figure SI-1.** Equilibrium chemical speciation of 10 mM Ni(II) in aqueous solution (A) without $SO_4^{2-}$; with (B) 0.05 mM; (C) 0.5 mM; (D) 5 mM; (E) 50 mM; and (F) 500 mM $SO_4^{2-}$ estimated using Visual MINTEQ.

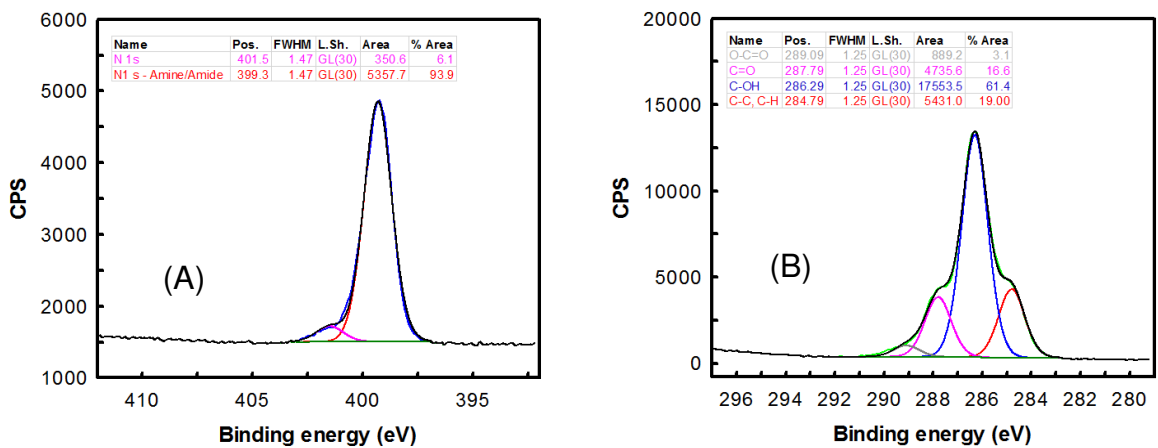

**Figure SI-2.** High-resolution XPS scan spectra over (A) N 1s and (B) C 1s of chitosan before Ni(II) adsorption.

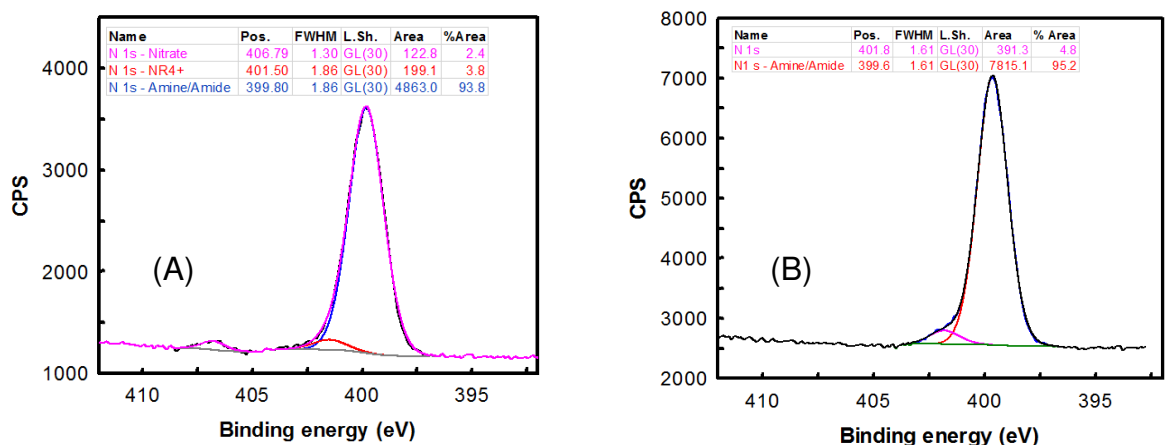

**Figure SI-3.** High-resolution XPS scan spectra over N 1s of chitosan after Ni(II) adsorption in the (A) absence and (B) presence of $SO_4^{2-}$.

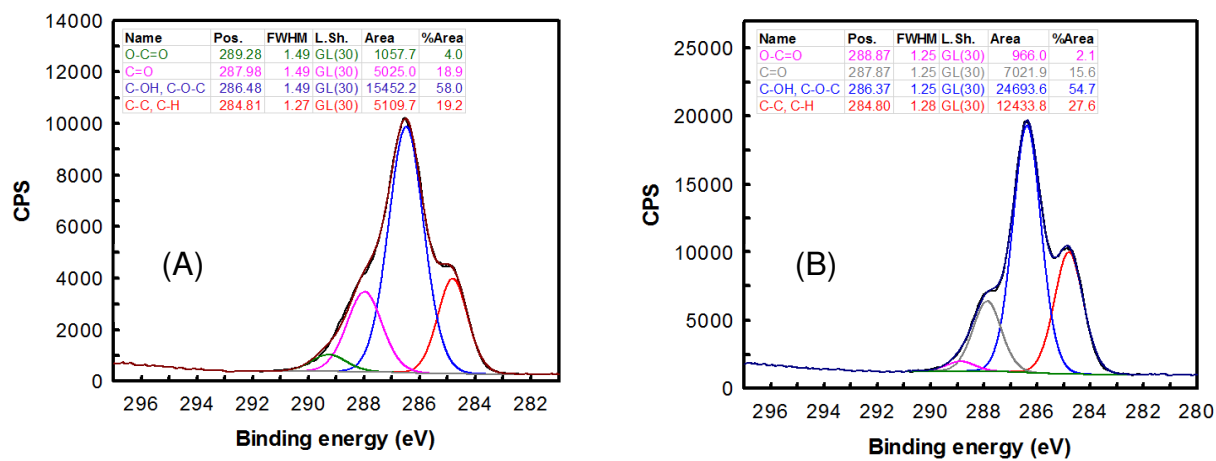

**Figure SI-4.** High-resolution XPS scan spectra over C 1s of chitosan after Ni(II) adsorption in the (A) absence and (B) presence of $SO_4^{2-}$.

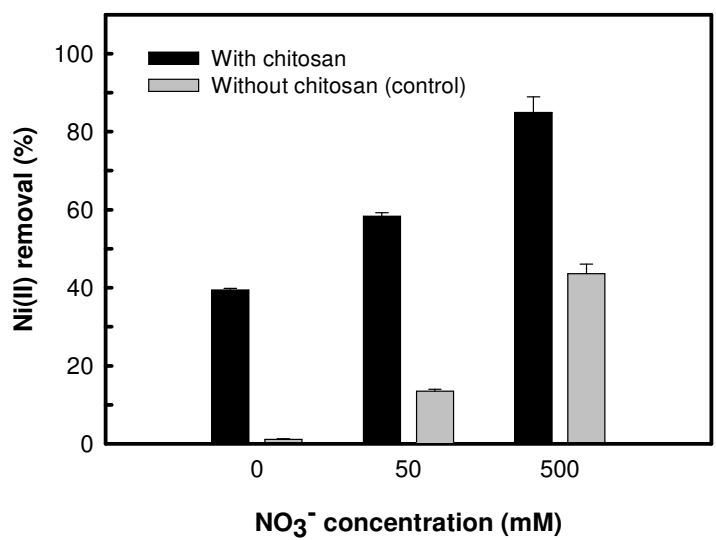

**Figure SI-5.** Effect of $NO_3^-$ concentration on Ni(II) removal efficiency of chitosan. Ni(II) removal is shown as the average value of the replicates (number of samples, N = 3) with error bars representing the standard deviation. Where not shown, error bars are too small. Experimental conditions: 10 mM initial Ni(II) concentration, 100 mL/g L/S ratio, initial pH 7, room temperature, 24 h contact time.

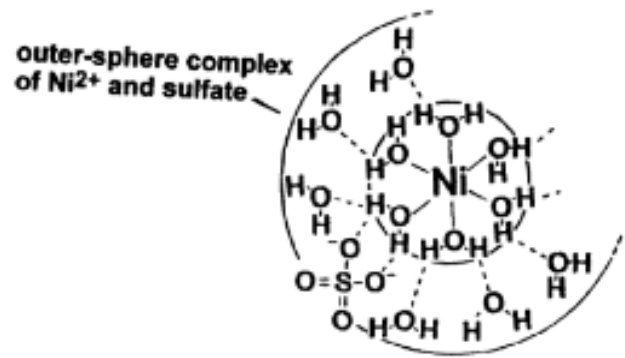

**Figure SI-6.** The formation of an outer-sphere complex of Ni(II) and $SO_4^{2-}$ (Martell and Hancock, 1996).

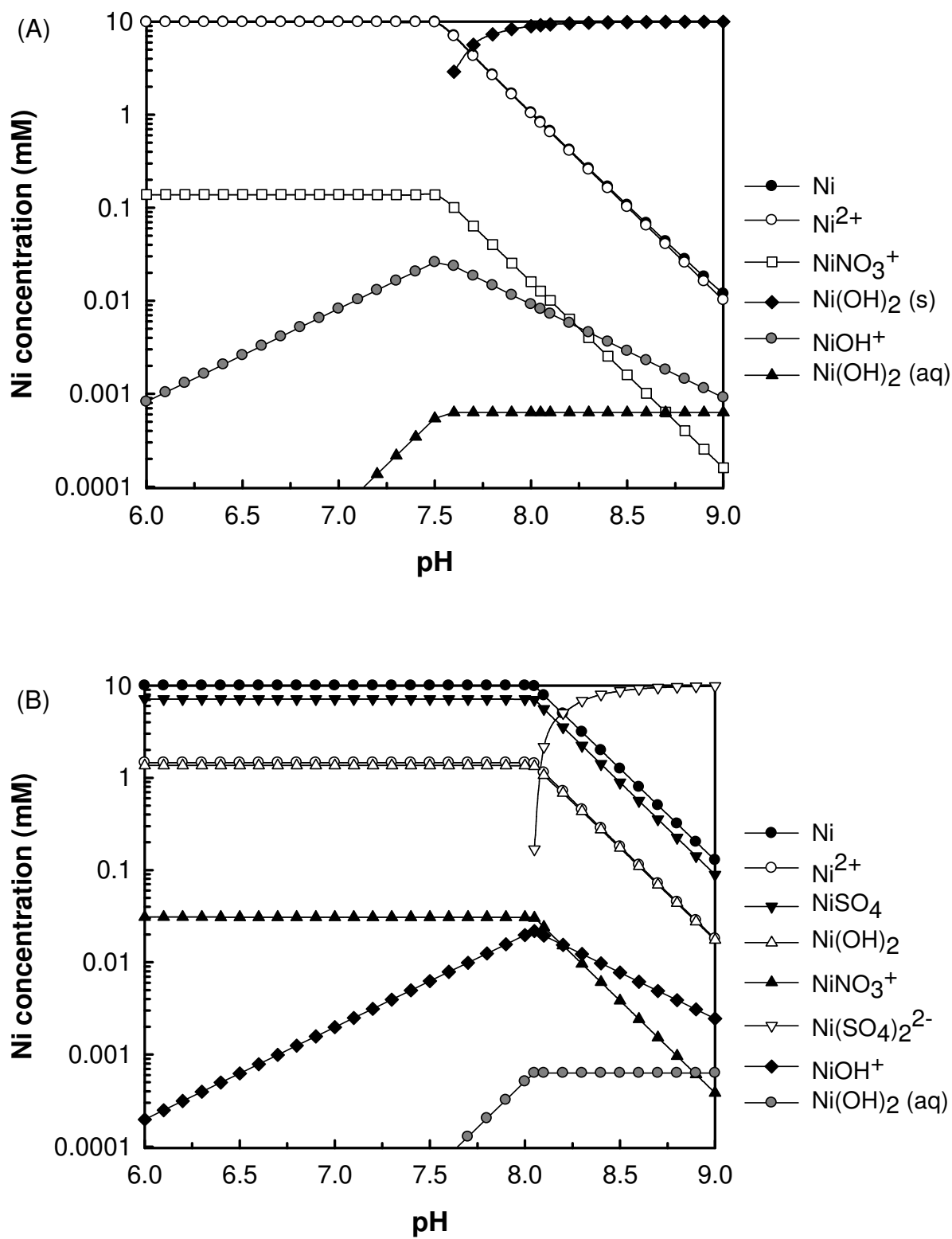

**Figure SI-7.** Equilibrium chemical speciation of 10 mM Ni(II) in aqueous solution (A) without $SO_4^{2-}$ and (B) with 500 mM $SO_4^{2-}$ estimated using PHREEQC.

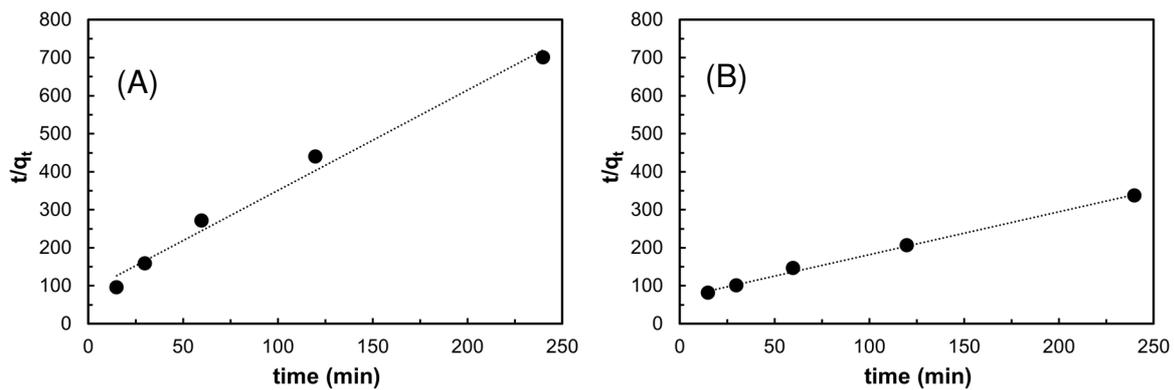

**Figure SI-8.** Pseudo -second order adsorption kinetics of Ni(II) on chitosan in a system (A) without $SO_4^{2-}$ and (B) with 500 mM $SO_4^{2-}$.